\documentclass[preprint,superscriptaddress]{revtex4}
\usepackage{graphics,epsfig}
\usepackage{epstopdf}
\usepackage{graphicx}
\usepackage{dcolumn}
\usepackage{amsmath}
\newcommand{\nn}{\nonumber}
\begin{document}
\title{Noncommutative geometry inspired rotating black string}
\author{Dharm Veer Singh}
\email{veerdsingh@gmail.com}
\affiliation{Centre for Theoretical Physics,
Jamia Millia Islamia, New Delhi 110025
India}
\author{Md Sabir Ali}
\email{alimd.sabir3@gmail.com}
\affiliation{Centre for Theoretical Physics,
Jamia Millia Islamia, New Delhi 110025
India}
\author{Sushant G. Ghosh}
\email{sgghosh@gmail.com,sghosh2@jmi.ac.in}
\affiliation{Centre for Theoretical Physics, and \\ Multidisciplinary Centre for Advanced Research and Studies (MCARS), Jamia Millia Islamia, New Delhi 110025, India}
\affiliation{Astrophysics and Cosmology Research Unit,
School of Mathematics, Statistics and Computer Science,
University of KwaZulu-Natal, Private Bag X54001,
Durban 4000, South Africa}
\begin{abstract}
	Noncommutativity is an idea dating back to the early times of Quantum Mechanics and that string theory induced noncommutative (NC) geometry which provides an effective framework to study short distance spacetime dynamics.  Also, string theory, a candidate for a consistent quantum theory of gravity, admits a variety of classical black hole solutions including  black strings.  In this paper, we study a NC geometry inspired rotating black string to cylindrical spacetime with a source given by a Gaussian distribution of mass. The resulting metric is a regular, i.e. curvature-singularity free, rotating black string, that in large $r$ limit interpolates Lemos \cite{lemos96} black string. Thermodynamical properties of the black strings are also investigated and exact expressions for the temperature, the entropy and the heat capacity are obtained. Owing to the NC correction in the solution, the thermodynamic quantities have been also modified and that the NC geometry inspired black string is always thermodynamically stable.
\end{abstract}
\maketitle
\section{\label{sec:level1}Introduction}
In general relativity, finding an exact solution to the Einstein field equations has been a subject of great interest, specially, the black hole solutions take an important place because thermodynamics, gravitational theory and quantum theory are connected via quantum black hole physics. The importance of spherical black holes has come from its role as the final state of complete gravitational collapse of a star, and hence it is useful to investigate if black holes with different topology may also emerge from gravitational collapse of some matter distribution. Many of these studies in the gravitational collapse were motivated by Thorne in hoop conjecture \cite{kst} that after collapse will yield a black hole only if a mass $ M $ is compressed into a region with circumference $C < 4 \pi M$ in all directions. If the hoop conjecture is true, naked singularities may form if collapse can yield $C > 4 \pi M$ in some direction. Thus, cylindrical matter will not form a black hole (or a black string) \cite{jps}. However, the hoop conjecture was imposed on spacetimes with a zero cosmological term, but when a negative cosmological constant is introduced, the spacetime will become asymptotically anti-de Sitter. Indeed, Lemos \cite{jps} has shown that cylindrical black holes are formed rather than naked singularity from gravitational collapse of a cylindrical matter distribution in an anti-de Sitter spacetime, violating in this way the hoop conjecture. In addition, string theory also gave a great variety of classical black hole solutions including black strings \cite{1,2,3,4,5,5.1,6,7,8,9,10,11,13,14,15}. Horowitz and Strominger \cite{14} showed that there does not exist static, cylindrically symmetric black hole (or a black string) solution with asymptotically flat spacetime in the transverse directions. However, Lemos \cite{lemos95,lemos951} constructed the cylindrical black hole in four dimensions which is asymptotically anti-de Sitter not only in the transverse direction but also in the string direction, and it has a corresponding three dimensional black hole solution \cite{lemos95}. The Lemos black string can be obtained starting from the Einstein-Hilbert action of the (3+1) dimensional gravity with a negative cosmological constant \cite{lemos95}
\begin{equation}
S=\frac{1}{16\pi G}\int d^4x\sqrt{-g}\,(R-2\Lambda) +S_{m},
\label{eqac}
\end{equation}
where $S_m$ is the action of matter field, $R$ is the curvature scalar, $g$ is the determinant of the metric, $\Lambda$ is the cosmological constant and $G$ is the gravitation constant. The equation of motion can be obtained by varying action (\ref{eqac}) with respect to the gravitational field $g_{ab}$ which yields
\begin{equation}
G_{ab}+3\Lambda g_{ab}=8\pi T_{ab}.
\label{eineq}
\end{equation}
Eq. (\ref{eineq}) admits the following black string solution \cite{lemos96}:
\begin{equation}
ds^2=-\Big(\alpha^2r^2-\frac{4M}{ r}\Big)\,d\bar{t}^2+\Big(\alpha^2r^2-\frac{4M}{ r}\Big)^{-1}\, dr^2+r^2\,d\bar{\phi}^2+ \alpha^2r^2\,dz^2,
\label{eqst}
\end{equation}
with $-\infty<\bar{t}<\infty,\quad 0\leq r\leq\infty,\quad 0\leq\bar{\phi}\leq2\pi,\quad -\infty<z<\infty$ and $M$ is ADM mass of black string. The constant $\alpha$ has the dimension of inverse of length which is related to cosmological constant by $\Lambda=-3\alpha^2$. The metric (\ref{eqst}) has three Killing vectors: $\partial/\partial z$ which corresponds to the translational symmetry along the axis, $\partial/\partial \bar{\phi}$ which has closed periodic trajectories around the axis and $\partial/\partial\bar{t}$ corresponding to the invariance under time translations. The rotating counterpart of the metric (\ref{eqst}) can be obtained by the following transformations \cite{lemos96}
\begin{equation}
\bar{t}=\lambda\,t-\frac{\omega}{\alpha^2} \phi,\qquad\qquad \bar{\phi}=\lambda \phi-\omega t.
\label{cot}
\end{equation}
and the metric of the rotating black string reads \cite{lemos96}:
\begin{eqnarray}
ds^2=&&-\Bigg[\left(\lambda^2-\frac{\omega^2}{\alpha^2}\right)\alpha^2r^2-\frac{4M\lambda^2}{ r}\Bigg]d{t}^2-\frac{8\omega \lambda M}{\alpha^3 r}\,d{t}\,d{\phi}+\frac{1}{\alpha^2 r^2-\frac{4M}{ r}} dr^2\nn\\&&+\Bigg[\left(\lambda^2-\frac{\omega^2}{\alpha^2}\right)r^2-\frac{4M\lambda^2}{ \alpha^4 r}\Bigg]d{\phi}^2+\alpha^2 r^2 dz^2,
\label{metric4}
\end{eqnarray}
where $\lambda$ and $\omega$ are constants. The metric (\ref{metric4}) is to be asymptotically anti-de Sitter, if we impose the condition $\lambda^2-\omega^2/\alpha^2=1$. It is useful to define the parameter $a$ and angular velocity $\Omega$ to study the solution and its causal structure, which are related to the angular momentum and angular velocity via
\begin{equation}
J=\frac{3}{2}aM\sqrt{1-\frac{a^2\alpha^2}{2}} \qquad \text{and}\qquad \Omega=\frac{a\alpha^2}{\sqrt{1-\frac{a^2\alpha^2}{2}}}
\end{equation}
The metric (\ref{metric4}) after rearrangements of various terms can be put in the following form
\begin{equation}
ds^2=-\Delta\Big(\lambda\,d{t}-\frac{\omega}{\alpha^2}\,d{\phi}\Big)^2+r^2(\lambda\, d{\phi}-\omega\, d{t})^2+\frac{dr^2}{\Delta}+\alpha^2 r^2 dz^2,
\label{metric3}
\end{equation}
where
\begin{eqnarray}
&&\Delta=\alpha^2r^2-\frac{4M}{ r}\left(1-\frac{3}{2}\alpha^2a^2\right)\nn\\
&& \lambda=\sqrt{\frac{1-\frac{1}{2}a^2\alpha^2}{1-\frac{3}{2}a^2\alpha^2}}\qquad {\text{and}} \qquad\omega=\frac{a\alpha^2}{\sqrt{1-\frac{3}{2}a^2\alpha^2}}.
\end{eqnarray}
The metric ({\ref{metric3}}) has many similarities with the metric on the equatorial plane for axis-symmetric rotating Kerr metric. The differences and similarities between these two solutions are given in \cite{lemos96,Cai96}, and we shall not discuss them here.
\section{Noncommutative Geometry Inspired Solution}
By now, it is clear that string theory influenced the NC geometry \cite{snyder,snyder1}, that provides an effective way to analyze the short distant spacetime dynamics. The NC spacetime was originally introduced by Snyder \cite{snyder, snyder1} to study the divergences in relativistic quantum field theory. On the other hand, the NC geometry appears naturally from the study of open string theories. In particular, the NC black holes are involved in the study of string theory and M-theory \cite{Haro,Haro1}. The noncommutativity eliminates the point-like structure in the form of smeared objects \cite{Myung:2008kp}. The NC geometry inspired Schwarzschild black hole is obtained by Nicolini et. al \cite{nicoli06}, which they also extended to the Reisnner-Nordstrom balck hole\cite{nicoli07}, generalized to Schwarzschild-Tangherilini spacetime by Rizzo \cite{Rizzo:2006zb}, to charged higher dimensions black holes \cite{nicoli09}, and then to the BTZ black holes \cite{Kim:2007nx} (see also Ref.~\cite{Nicolini:2006}, for a review of NC geometry inspired black holes). It turns out that the NC geometry inspired Schwarzschild black hole smoothly interpolates between a de-Sitter core at short distance and the ordinary Schwarzschild black hole at large $r$ \cite{nicoli06}. Our aim is to find an NC geometry inspired black string i.e. to find the NC counterpart of the solutions (\ref{eqst}) and (\ref{metric3}) and discuss their thermodynamical properties. The NC theories are built on spaces whose coordinate do not commute and therefore one cannot localize their points with an arbitrary precision.
In the NC geometry inspired black hole spacetime, the point-like object is replaced by the smeared object, whose mass density is described by a Gaussian distribution of some minimal width. The modified black hole due to noncommutativity of spacetime does not allow the black hole to decay beyond a minimal mass. Then, the evaporation processes terminate when the black hole reaches a Planck scale remnant at zero temperature, which does not diverse at all. In NC spacetime, a description of the point mass is not possible due to the fuzziness of spacetime. The fuzziness can be introduced by the following relation:
\begin{equation}
[x^{\mu},x^{\nu}]=i\theta^{\mu\nu},
\end{equation}
where $\theta^{\mu\nu}$ is an antisymmetric matrix, which determines the fundamental cell discretization of spacetime much in the same way as the ${\hbar}$ discretizes the phase space. Clearly, at a distance $\sqrt{\theta}$, the classical concept of smooth spacetime is no longer valid, where $\sqrt{\theta}$ is close to the Planck length.
The expression of the mass density which is smeared into a minimal width Gaussian profile for cylindrically symmetric spacetime is given as \cite{Myung:2008kp}
\begin{equation}
\rho_{\theta}=\frac{M}{ 4(\pi\;\theta)^{3/2}}e^{-r^2/4\theta},
\label{eq:nc}
\end{equation}
which is a matter source, and total mass $M$ is diffused in the region of size $\sqrt{\theta}$.
In order to completely define the energy – momentum tensor, we rely on the covariant conservation condition $T_{\mu;\nu}^{\nu} = 0$, which for the cylindrically symmetric metric (\ref{eqst}) reads
\begin{eqnarray}
\partial_rT^r_r&&=-\frac{1}{2}g^{tt}\partial_rg_{tt}(T^r_r-T^t_t),\nn\\
&&=g^{zz}\partial_rg_{z z}(T^r_r-T^{z}_{z}).
\label{em1}
\end{eqnarray}
For a black string solution, we require that $g_{tt}=-1/g_{rr}$ \cite{nicoli06}. Then Eq. (\ref{em1}) leads to the following relations
\begin{eqnarray}
&&p_{r}=-\rho_{\theta}=\frac{M}{4(\pi\theta)^{3/2}}e^{-r^2/4\theta},\nn\\
&&p_{\phi}=p_z=-\rho_\theta-\frac{r}{2}\partial_r \rho_\theta.
\label{em2}
\end{eqnarray}
Thus, the energy momentum tensor for the anisotropic perfect fluid is given as $T^\mu_\nu=\text{diag}[-\rho_\theta,p_r,p_\phi,p_z]$ and is completely known with $\rho_\theta$, the energy density of the anisotropic perfect fluid, whereas, $p_r$ and $p_\phi$ or $p_z$ respectively are the radial and the tangential pressure. On solving the Einstein's equations (\ref{eineq}) for the energy-momentum tensor (\ref{em2}), one obtains
\begin{equation}
ds^2=-\Big[\alpha^2r^2-\frac{8M}{\sqrt{\pi} r}\gamma\Big(\frac{3}{2} ,\frac{r^2}{4\theta}\,\Big)\Big]\,dt^2+\Big[\alpha^2r^2-\frac{8M}{\sqrt{\pi} r}\gamma\Big(\frac{3}{2} ,\frac{r^2}{4\theta}\,\Big)\Big]^{-1}\, dr^2+r^2\,d\phi^2+ \alpha^2r^2\,dz^2,
\label{ncmet1}
\end{equation}
where
\begin{equation}
\gamma\Big(\frac{b}{a},\frac{r^2}{4 \theta}\Big) = \int_0^{{r^2}/{4 \theta}} u^{b/a} e^{-u} \frac{du}{u},
\end{equation}
is the lower incomplete gamma function.
The surface area per unit height on $z$-axis is $A_+=2 \pi (r \sqrt{g_{\phi\;\phi}})_{r=r_+}$
and hence the mass parameter $m_\theta(r)$ is related to mass density (\ref{eq:nc}) by
\begin{equation}
m_{\theta}(r)=\int_{0}^{r} 2\pi r^{\prime 2}\rho_{\theta}(r^{\prime}) dr^{\prime} =\frac{2M}{\sqrt{\pi}}\gamma \left(\frac{3}{2},\frac{r^2}{4\theta}\right),
\label{masstheta}
\end{equation}
where $\sqrt{\theta}$ is the minimal width of the Gaussian profile given in (\ref{eq:nc}). The solution (\ref{ncmet1}) can be also obtained by replacing $M$ by $m_{\theta}(r)$ in Eq. (\ref{eqst}) \cite{51, 50}. In the limit $ r/\sqrt{\theta} \to \infty$, Eq. (\ref{ncmet1}) reduces to the Lemos \cite{lemos96} black string solution $(\ref{eqst})$. Thus, we see that instead of being concentrated at a point, the sourse is spreading throughout a region of linear size $\sqrt{\theta}$ because of the noncommutativity. We note that the following properties of matter density: near the origin $r<< \sqrt{\theta}$, $d\rho_{\theta}/dr\simeq 0\rightarrow\rho_{\theta}\simeq \rho(0)$, deviation from the origin $r\geq \sqrt{\theta}$, $d\rho_{\theta}/dr\simeq 0\rightarrow\rho_{\theta}\simeq 0$ and asymptotically we get $\rho_{\theta}(r)=0$.
The horizon of the NC geometry inspired black string is located at
\begin{equation}
r_+^3=\frac{8M}{\sqrt{\pi}\alpha^2}\gamma\Big(\frac{3}{2},\frac{r_+^2}{4\theta}\Big).
\label{hor1}
\end{equation}
\begin{figure}[h]
\centering
\includegraphics[width=0.8\textwidth]{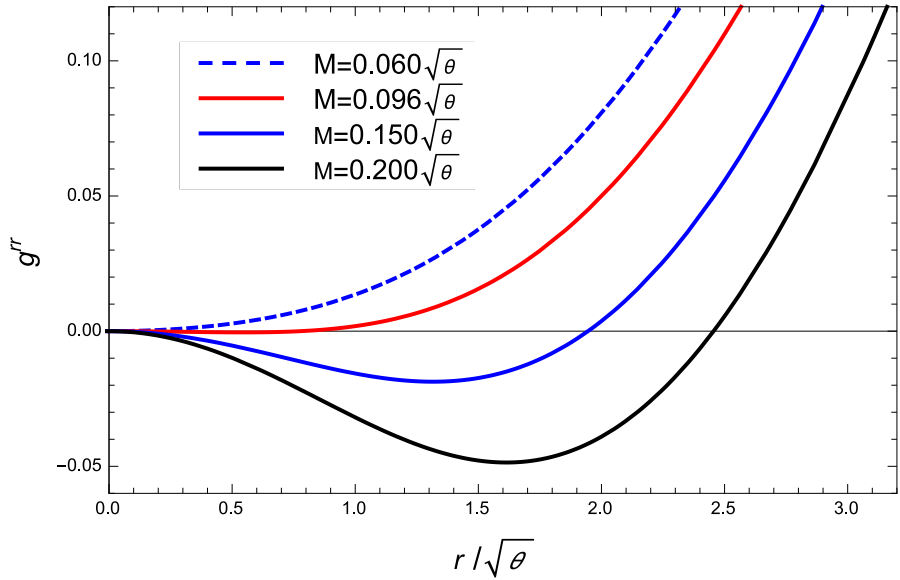}
\caption{ The plot of metric function $g^{rr}$ vs $r/\sqrt{\theta}$ with fixed value of $\alpha{\sqrt{\theta}}=0.57735$ for NC geometry inspired black string.}
\label{fig:hor}
\end{figure}

\noindent However, Eq. (\ref{hor1}) can not be solved analytically, and hence we plot them in Fig. \ref{fig:hor} with different values of mass parameter. The horizon radius increases with an increase in the mass parameter. The large distance limit $r/\sqrt{\theta}\to\infty$ gives the black string event horizon located at $r_+^3=4M/\alpha^2$. Thus the metric (\ref{ncmet1}) describes a NC geometry inspired black string and it is useful to see how the noncommutativity affects the thermodynamical properties.

{In the large radius $r^2/4\theta\gg 1$, the NC effect can be neglected. But at the short distance $r^2/4\theta$ $\approx O(1)$, one expects important changes. It turns out that NC makes the space-time regular which can be seen by the behavior of curvature scalar ($R$), Ricci square ($R_{ab}R^{ab}$) and Kretschmann invariant ($R_{abcd}R^{abcd}$) of the NC geometry inspired black string (\ref{ncmet1}). These invariants are calculated at the origin, which read}
\begin{eqnarray}
&&\lim_{r\to 0} R =-3\alpha^2+\frac{2M}{\pi^{1/2}\theta^{3/2}},\nn\\
&&\lim_{r\to 0}R_{ab}R^{ab}=\frac{324\alpha^8-\frac{864M\alpha^4}{\pi^{1/2}\theta^{3/2}}\left(\alpha^2-\frac{M}{\pi^{1/2}\theta^{3/2}}\right)-\frac{64M^3}{\pi^{3/2}\theta^{9/2}}\left(6\alpha^2-\frac{M}{\pi^{1/2}\theta^{3/2}}\right)}{9\alpha^4-\frac{4M}{\pi^{1/2}\theta^{3/2}}\left(3\alpha^2-\frac{M}{\pi^{1/2}\theta^{3/2}}\right)},\nn\\
&&\lim_{r\to 0} R_{abcd}R^{abcd}=24\alpha^4-\frac{32M}{\pi^{1/2}\theta^{3/2}}\left(\alpha^2-\frac{M}{3\pi^{1/2}\theta^{3/2}}\right).
\label{inv}
\end{eqnarray}
We see from the above expressions of invariants (\ref{inv}) that they are finite and regular everywhere including at the origin for the non-zero values of ${\theta}$. These invariants are also regular even if we take $M=0$ and hence NC black string are regular solution.
\section{Thermodynamics}
\begin{figure}[h]
\centering
\includegraphics[width=0.8\textwidth]{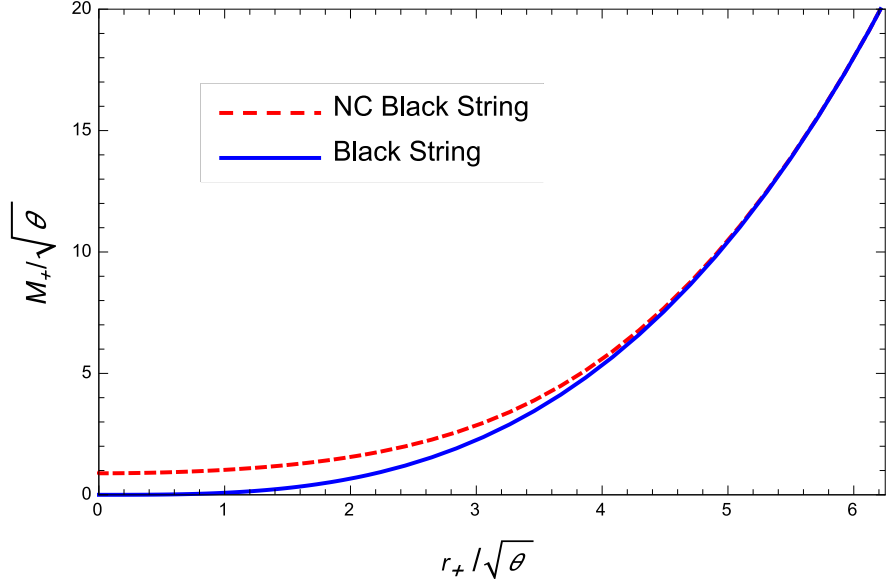}
\caption{The plot of mass $M_+\sqrt{\theta}$ vs horizon radius $r_+/\sqrt{\theta}$ with fixed value of $\alpha{\sqrt{\theta}}=0.57735$ for NC inspired black string.}
\label{fig:mpbs}
\end{figure}

In the previous section, we showed that NC inspired black string  (\ref{ncmet1}) admits two horizons.  Next, we calculate exact thermodynamic quantities in terms of horizon radius ($r_+$) associated with the NC geometry inspired black string (\ref{ncmet1}). From the equation $g_{tt}(r_+)=0$, the mass of the NC geometry inspired black string can be expressed as
\begin{equation}
M_+=\frac{\sqrt{\pi}\,r_{+}^3\,\alpha^2}{8\,\gamma\big(\frac{3}{2},\frac{r_{+}^2}{4\theta}\big) }.
\label{eqmp}
\end{equation}
\noindent{In the large radius limit, $r_+/\sqrt{\theta}\to \infty$, the NC effect vanishes and we get the mass of the black string, $M_+=\alpha^2 r_+^3/4$ \cite{lemos95, lemos951, lemos96}. However, when $r_+\sim \sqrt{\theta}$ the noncommutativity becomes important \cite{nicoli07,nicoli09} as shown in Fig. \ref{fig:mpbs}. In NC case the mass does not vanish, rather it gives the finite value $(M_+={3\sqrt{\pi \theta}\alpha^2\theta^{2}}/{2})$, even when the horizon radius shrinks to zero (cf. Fig. \ref{fig:mpbs} ). This minimal mass of NC black string corresponds to the Planck scale remnant beyond which no further decay of NC black string is possible. But in the standared black string solution (\ref{eqst}), mass term vanishes as horizon radius approaches to zero. The mass of two solutions (\ref{eqst}) and (\ref{metric3}) coincides at $r_+=6.21\sqrt{\theta}$ (cf. Fig. \ref{fig:mpbs}).}

\begin{figure}[h]
\centering
\includegraphics[width=0.8\textwidth]{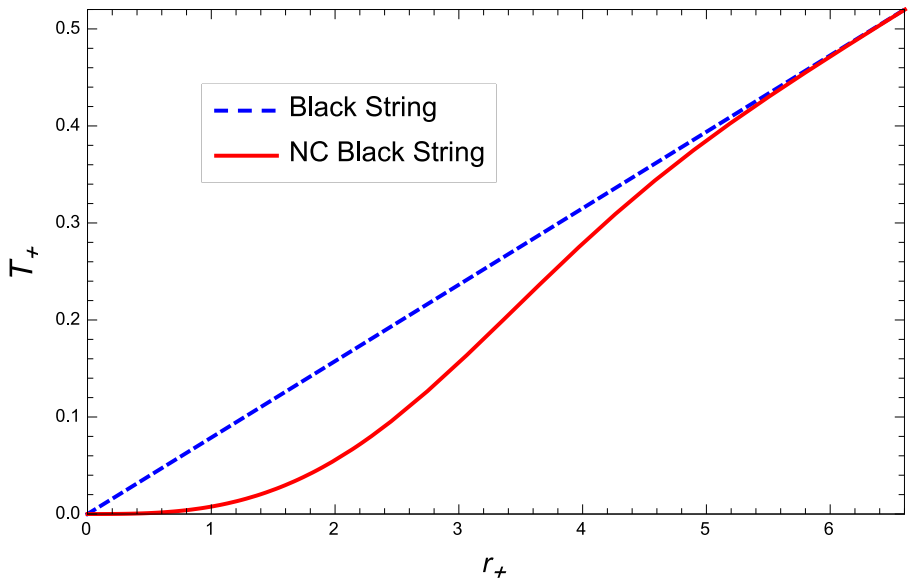}
\caption{\label{fig:tpbs}The Hawking temperature $(T_+)$ vs  horizon radius $(r_+)$ with fixed value of $\alpha=0.57735$ in $\sqrt{\theta}$ units for NC geometry inspired black string.}
\end{figure}

\noindent{The Hawking temperature associated with the black string is given by $T=\kappa/2\pi$, where $\kappa$ is the surface gravity and it is defined as}
\begin{equation}
\kappa^2=-\frac{1}{2}\nabla_{\mu}\xi_{\nu}\nabla^{\mu}\xi^{\nu},
\label{kill1}
\end{equation}
where $\xi{^\mu}=\partial_t$ is the Killing vector of the metric (\ref{ncmet1}).
On using (\ref{ncmet1}) the temperature of the NC geometry inspired black string can be expressed as
\begin{equation}
T_+=\frac{1}{4\pi}\partial/\partial r\sqrt{-g^{rr}g_{tt}}|_{r=r_+}=\frac{1}{4\pi}{\alpha^2 r_{+}}\Bigg[3-r_+\,\frac{\gamma'\left(\frac{3}{2},\,\frac{r_{+}^2}{4\theta}\right)}{\gamma\left(\frac{3}{2},\,\frac{r_{+}^2}{4\theta}\right)}\Bigg].
\label{eqtp}
\end{equation}
Eq. (\ref{eqtp}) gives the NC geometry inspired black string temperature and in large radius limit, $r/\sqrt{\theta}\to\infty$, one recovers the temperature of the commutative black string :
\begin{equation}
\label{eqtc}
T_+=\frac{3\alpha^2 r_+}{4\pi}.
\end{equation}
\noindent The temperature is a monotonically increasing function of horizon radius. Therefore at the initial stage of black hole evaporation, the temperature decreases as the horizon radius. In NC case the temperature does not increase linearly as it does in the case of black string. The spacings between the two temperatures given by (\ref{eqtp}) and (\ref{eqtc}) decreases as we go from smaller to larger horizon radii and becomes maximum at $r_+=2.45\sqrt{\theta}$ corresponding to temperature $T_+=0.0945\sqrt{\theta}$ (see in Fig. \ref{fig:tpbs}). As we can see clearly, the graph of two temperatures deviates significantly upto few order of standard deviation in the noncommutative limit of spacetime geometry and they coincide at $r_+=6.55\sqrt{\theta}$. The zero temperature corresponds to the termination of evaporation process of the black string.

Since black string behaves as a thermodynamical system, the expression of the entropy can be obtained from the first law of thermodynamics \cite{cai02,c06,ghosh14},
\begin{equation}
dM_+=T_+ dS_+,
\end{equation}
thus on integrating the first law, we have
\begin{eqnarray}
S_+&&=\int \frac{1}{T_+}\,dM_+=\int_0^{r_+}\,\frac{1}{T_+}\Big(\frac{\partial M_+}{\partial r_+}\Big)dr_+,
\label{eqS}
\end{eqnarray}
on substituting Eq. (\ref{eqmp}) and Eq. (\ref{eqtp}) into Eq. (\ref{eqS}), we find the entropy of the NC geometry inspired black string as
\begin{equation}
S_+=\frac{\pi^{3/2}}{2}\,\int_0^{r_+}\frac{ r_+}{\gamma\big(\frac{3}{2},\frac{r^2_+}{4\theta}\big)}\,dr_+.
\label{eqS1}
\end{equation}
The entropy of the NC geometry inspired black string does not satisfy the area law. However, in the large radius limit the entropy obeys the Bekenstein area law, $S=\pi r_+{^2}/2=A/4$, where $A=2\pi r_+^2$ is the area of the black string \cite{Cai96}.

To analyze  the thermodynamical stability of NC geometry inspired black string (\ref{ncmet1}),  we evaluate the heat capacity. The heat capacity of the black string is defined as
\begin{eqnarray}
C_+=\frac{\partial M_+}{\partial T_+}=\Big(\frac{\partial M_+}{\partial r_+}\Big)\slash \Big(\frac{\partial T_+}{\partial r_+}\Big).
\label{eqcp}
\end{eqnarray}
Substituting (\ref{eqmp}) and (\ref{eqtp}) into (\ref{eqcp}), the heat capacity of the NC geometry inspired black string reads
\begin{eqnarray}
C_+=\frac{\pi^{3/2} r_{+}^2\Bigg[3-\frac{r_+\gamma'\left(\frac{3}{2},\,\frac{r_{+}^2}{4\,\theta}\right)}{\gamma\left(\frac{3}{2},\,\frac{r_{+}^2}{4\,\theta}\right)}\Bigg]}{2\Bigg[3\gamma\left(\frac{3}{2},\,\frac{r_{+}^2}{4\,\theta}\right) -r_+\left[2\gamma'\left(\frac{3}{2},\,\frac{r_{+}^2}{4\,\theta}\right)+r_+\gamma''\left(\frac{3}{2},\,\frac{r_{+}^2}{4\,\theta}\right)\right]+r_+^2\frac{\gamma'^2\left(\frac{3}{2},\,\frac{r_{+}^2}{4\,\theta}\right)}{\gamma\left(\frac{3}{2},\,\frac{r_{+}^2}{4\,\theta}\right)}\Bigg]}.
\label{cap1}
\end{eqnarray}
In the limit when $r_+/\sqrt{\theta} \to \infty $, one finds that
\begin{equation}
C_+=\pi r_+{^2},
\label{cap2.1}
\end{equation}
\begin{figure}[h]
\begin{tabular}{c c c c}
\includegraphics[width=0.8\linewidth]{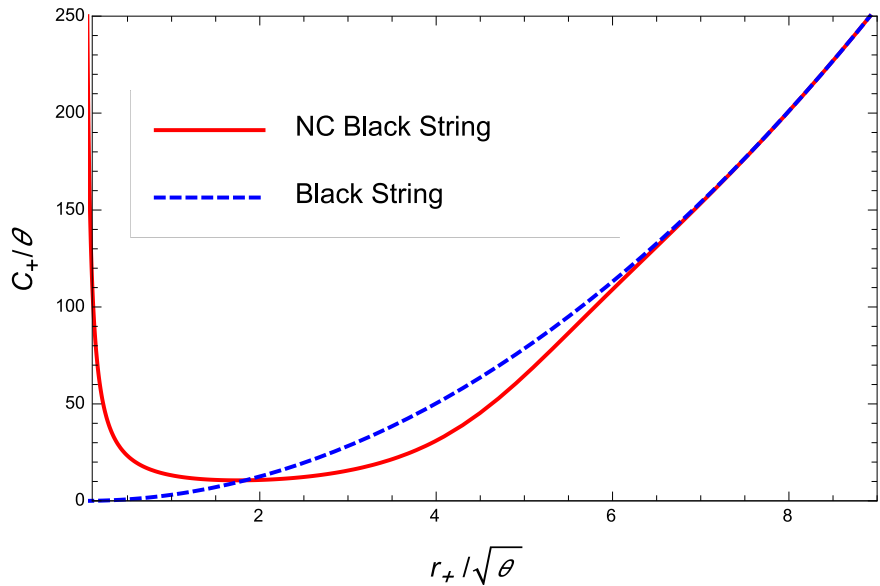}
\end{tabular}
\caption{\label{fig:cbs} The specific heat $(C_+/\theta)$ vs  horizon radius $(r_+/\sqrt{\theta})$ for the NC geometry inspired black string. The solid and dashed line show the heat capacity of NC geometry inspired black string and black string respectively.}
\end{figure}

\noindent which is the specific heat of commutative black string (\ref{eqst}). For large $r_+$ two black string solutions (\ref{eqst}) and (\ref{ncmet1}) have the nearly same heat capacity as shown in Fig. \ref{fig:cbs}. The thermodynamical stability of the system is related to the sign of the heat capacity. If it is positive ($C_+>0$), then the black string is thermodynamically stable, when it is negative ($C_+<0$), the black string is said to be thermodynamically unstable.
For the standard black string case, we have parabolic variation of the horizon radius $r_+$ with the heat capacity. But when noncommutativity encounters, scenario becomes very important for the horizon radius up to few order of standard deviation. Thus, inclusion of the NC factor into the black string solution has direct impact on heat capacity at short distances. In the standard black string solution heat capacity vanishes as horizon radius shrinks to zero. But in case of NC black string even if the horizon radius approaches to zero, the heat capacity never meets the origin, rather it gives a finite value. As the heat capacity is always positive, both the NC geometry inspired black string and standard black string are always thermodynamically stable (cf. Fig. \ref{fig:cbs}) and no further phase transition is possible.
\section{ Rotating Black String Solution}
The metric of rotating counterpart of NC geometry inspired black string (\ref{ncmet1}) reads:
\begin{equation}
ds^2=-\Delta\Big(\lambda\,dt-\frac{\omega}{a^2}\,d\phi\Big)^2+r^2(\lambda d\phi-\omega dt)^2+\frac{dr^2}{\Delta}+\alpha^2 r^2 dz^2,
\label{2}
\end{equation}
with
\begin{equation}
\Delta=\alpha^2r^2- \frac{8M}{\sqrt{\pi} r}\gamma\Big(\frac{3}{2},\frac{r^2}{4\theta}\Big)(1-\frac{3}{2}\alpha^2a^2),
\label{3}
\end{equation}
where $a$ is the rotation parameter. The event horizon of NC geometry inspired rotating black string is obtained by $g^2_{t\phi}-g_{tt}g_{\phi\phi}=0$, which in turn yields $\Delta=0$. From the expression (\ref{3}) we see that the rotation parameter should have the range $0\leq\alpha^2 a^2 < {2}/{3}$ \cite{lemos96} for the horizon radius to have the real values. Thus, the horizon of (\ref{2}) is found to be
\begin{equation}
r_+^3=\frac{8M}{\sqrt{\pi}\alpha^2}(1-\frac{3}{2}\alpha^2a^2)\gamma\Big(\frac{3}{2},\frac{r_+^2}{4\theta}\Big).
\label{4}
\end{equation}
\begin{figure}[h]
\centering
\includegraphics[width=0.8\textwidth]{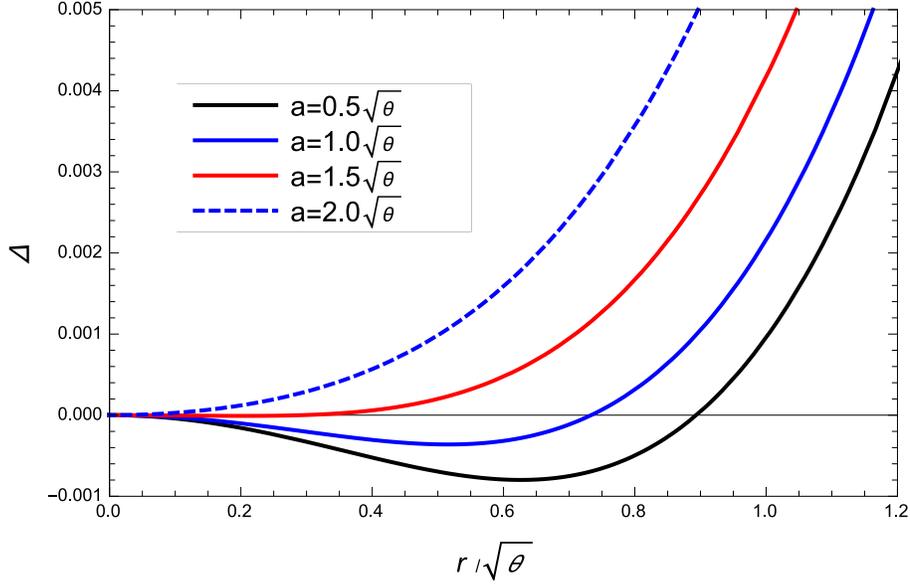}
\caption{The plot of $\Delta$ vs $r/\sqrt{\theta}$ with the different value of rotation parameter $a$ for NC geometry inspired rotating black string with fixed value of $M=0.10{\sqrt{\theta}}$ and $\alpha\sqrt{\theta}=0.57735$.}
\label{fig:hor2.2}
\end{figure}
\noindent The Eq. (\ref{4}) can't be solved analytically and hence it is shown in Fig. \ref{fig:hor2.2} for different values of rotation parameter $a$ with the fixed value of $M$ in terms of $\sqrt{\theta}$ units. The horizon radius increases with the decrease in rotation parameter $a$. In the large distance limit $r/\sqrt{\theta}\to 0$, Eq. (\ref{4}) gives the rotating black string horizon $ r_+^3={4M}/{\alpha^2}(1-{3}/{2}\alpha^2a^2)$ and $a\to 0 $ the horizon is located at $r_+^3={4M}/{\alpha^2}$ corresponding to non-rotating case. Thus, the metric (\ref{2}) together with Eq. (\ref{3}) describes the NC geometry inspired rotating black string and is useful to see how the noncommutativity affects the thermodynamical properties including stability.
Here, we calculate thermodynamical quantities associated with the NC geometry inspired rotating black string. From Eq. (\ref{3}) the mass of the NC geometry inspired rotating black string can be expressed in term of $r_+$ as
\begin{equation}
M_+=\frac{\sqrt{\pi}r^3_+\alpha^2}{8(1-\frac{3}{2}a^2\alpha^2)\gamma\Big(\frac{3}{2},\frac{r^2_+}{4\theta}\Big)}.
\label{eqmr}
\end{equation}
The large radius limit, $r_+/\sqrt{\theta}\to \infty$, leads to
\begin{equation}
M_+=\frac{r^3_+\alpha^2}{4(1-\frac{3a^2\alpha^2}{2})}.
\end{equation}

\begin{figure}[h]
\centering
\begin{tabular}{c c c c}
\includegraphics[width=0.55\textwidth]{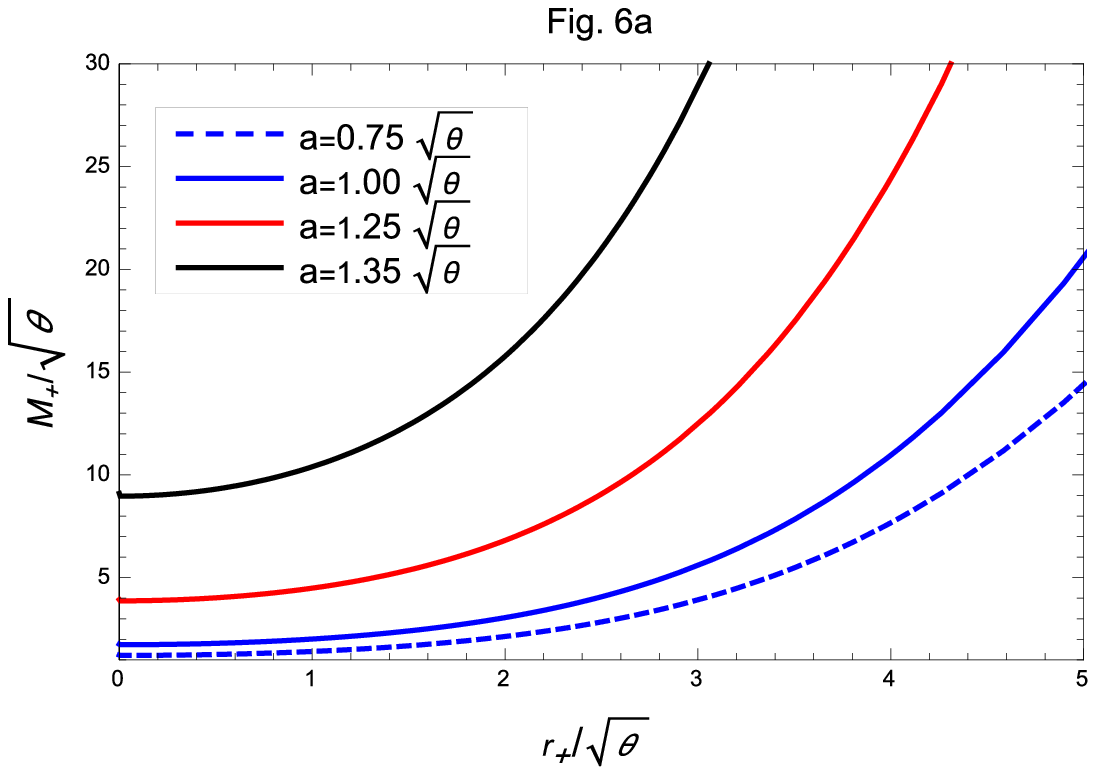}
\includegraphics[width=0.55\textwidth]{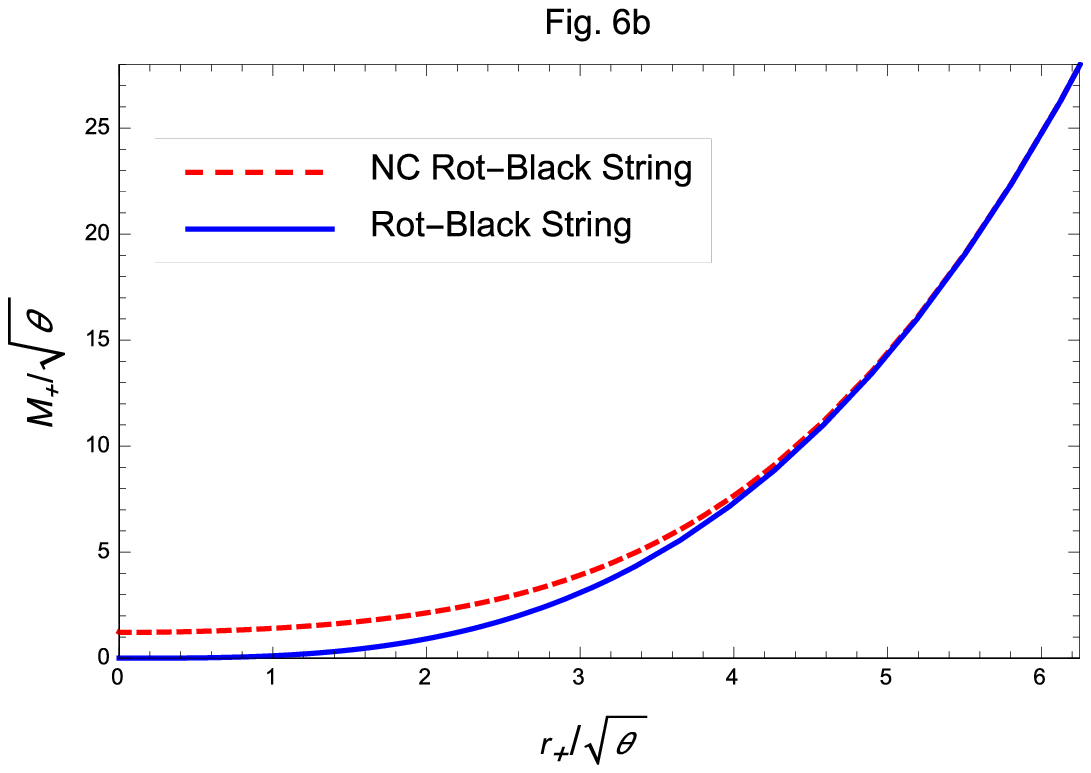}
\end{tabular}
\caption{The plot of mass vs   $r_+/\sqrt{\theta}$ with the different value of rotation parameter $a$ and fixed value of $\alpha\sqrt{\theta}=0.57735$. In second figure the solid and dashed lines show the mass of rotating black string and NC geometry inspired rotating black string for fixed value of rotation parameter $a=0.75\sqrt{\theta}$.}
\label{CS312}
\end{figure}

\noindent When $a \to 0$ in Eq. (\ref{eqmr}), one gets the mass of NC geometry inspired black string (\ref{eqst}). We observe that the mass of the NC geometry inspired black string increases with horizon radius and also increases with the rotation parameter $a$ (cf. Fig. \ref{CS312}a). In NC case the mass does not vanish, rather it gives a finite value, even when the horizon radius shrinks to zero (cf. Fig. \ref{CS312}b). But in the standared black string solution (\ref{eqst}) mass vanishes as horizon radius approaches to zero. The mass of NC geometry inspired black string coincides with that of the standard black string at $r_+=6.21\sqrt{\theta}$ with $a=0.75\sqrt{\theta}$ (cf. Fig. \ref{CS312}b). The rotating black string has larger mass as compared to non-rotating case for the same values of horizon radius.

{The temperature of the rotating black string is obtained through Eq. (\ref{kill1}). The rotating black string has two Killing vectors: ${\partial }/{\partial t}$ and ${\partial }/{\partial \phi}$, which amount to time translational and rotational symmetry. Thus, the Killing field is expressed as $ \xi^{\mu}={\partial }/{\partial t}+\Omega{\partial }/{\partial \phi}$.
Eq. (\ref{kill1}) together with the solution (\ref{2}) gives the expression of the temperature of NC geometry inspired rotating black string as

\begin{equation}
T_+=\frac{1}{2\pi}\frac{{\alpha^2 r_{+}}}{(2-{3a^2\alpha^2})}{\left[1+\frac{a\alpha}{\sqrt{1-\frac
{a^2\alpha^2}{2}}}\right]}\left[3-r_+\,\frac{\gamma'\left(\frac{3}{2},\,\frac{r_{+}^2}{4\theta}\right)}{\gamma\left(\frac{3}{2},\,\frac{r_{+}^2}{4\theta}\right)}\right],
\label{eqtr}
\end{equation}
we can find the temperature of rotating black string in the limit $r/\sqrt{\theta}\to \infty$ as
\begin{equation}
T_+=\frac{1}{2\pi}\frac{{3\alpha^2 r_{+}}}{(2-{3a^2\alpha^2})}{\left[1+\frac{a\alpha}{\sqrt{1-\frac
{a^2\alpha^2}{2}}}\right]}.
\label{eqtr1}
\end{equation}

\begin{figure}[h]
\centering
\begin{tabular}{c c c c}
\includegraphics[width=0.55\textwidth]{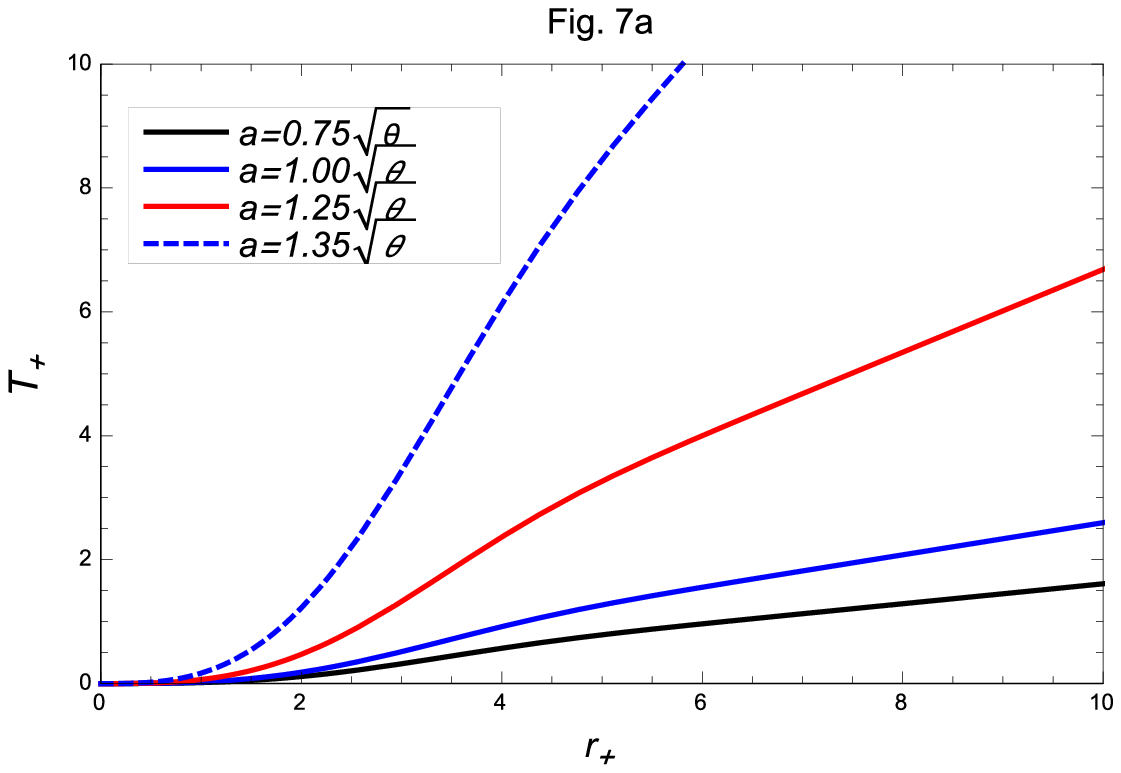}
\includegraphics[width=0.55\textwidth]{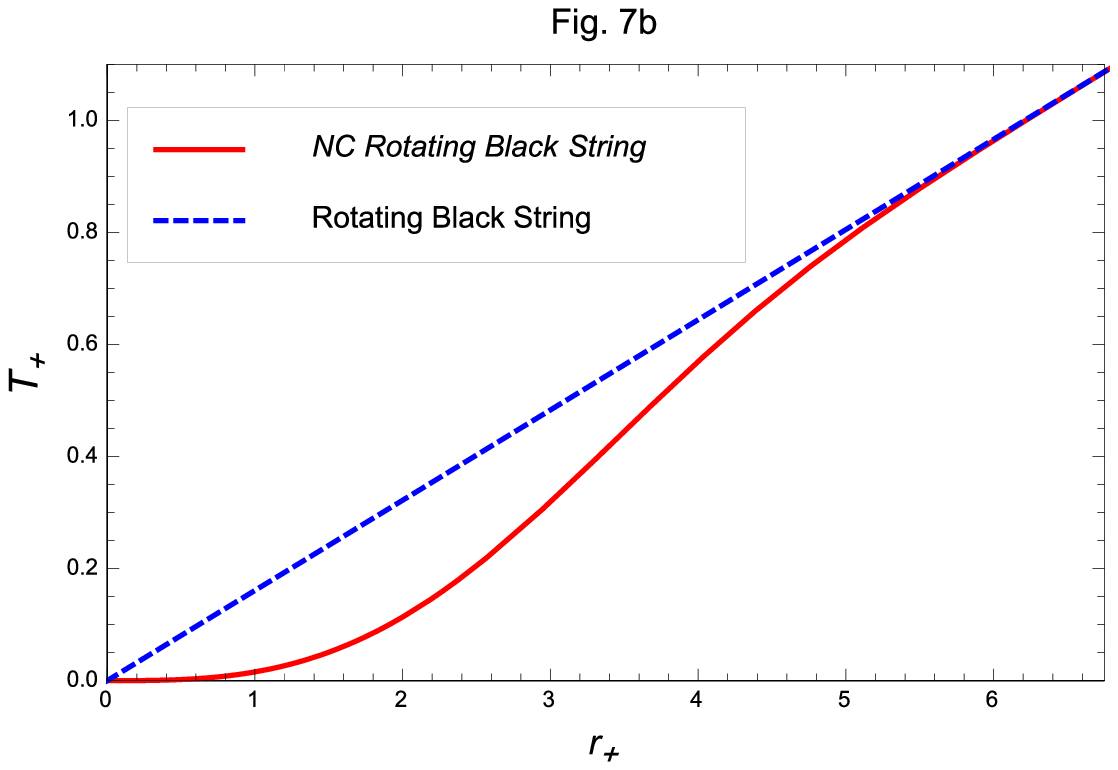}
\end{tabular}
\caption{The Hawking temperature $(T_+)$ vs horizon radius $r_+$ for the different value of rotation parameter $a$ in $\sqrt{\theta}$ units with fixed value of $\alpha\sqrt{\theta}=0.57735$ and in second graph the solid and dashed lines show the temperature of NC geometry inspired rotating black string and rotating black string with fixed value of $a=0.75\sqrt{\theta}$.}
\label{fig:mrbs1}
\end{figure}
\noindent This temperature (\ref{eqtr1}) is also obtained by Lemos but they calculated it in terms of mass after Euclideanizing the rotating black string \cite{lemos95}. The temperature profile looks similar to the non-rotating case. The temperature shifts toward the smaller values of horizon radius $r_+$ with increase in the rotation parameter $a$. Having a positive defined temperature $a$ cannot exceed the interval $0<\alpha^2 a^2<{2}/{3}$. As $\alpha^2a^2$ approaches to 2/3, the temperature diverges. In the region $r_+\simeq\sqrt{\theta}$,  the temperature deviates from the standard result (cf. Fig. \ref{fig:mrbs1}a). The temperature of NC geometry inspired rotating black string (\ref{eqtr}) coincides with rotating black string (\ref{eqtr1}) at $r_+=6.55\sqrt{\theta}$ (cf. Fig. \ref{fig:mrbs1}b).

Now, the entropy of the NC geometry inspired rotating black string is determined by using the first law of thermodynamics
\begin{equation}
dM_+ = T_+ dS_+ +\Omega_+\,d J
\end{equation}
we obtain the expression of entropy for NC geometry inspired rotating black string,
\begin{eqnarray}
S_+&&=\int_0^{r_+}\,\frac{1}{T_+}\Big(\frac{\partial M_+}{\partial r_+}dr_+-\Omega_+\frac{\partial J}{\partial r_+}dr_+\Big),\nn\\
&&=\int_0^{r_+}\frac{1}{T_+} \Big(\frac{\partial M_+}{\partial r_+}dr_+ -\frac{3}{2}a^2\alpha^2\frac{\partial M_+}{\partial r_+}dr_+\Big),
\label{eqsr}
\end{eqnarray}
substituting (\ref{eqmr}) and (\ref{eqtr}) into (\ref{eqsr}), we find the entropy of NC geometry inspired rotating black string as
\begin{equation}
S_+=\frac{\pi^{3/2}}{2\zeta} \int \frac{r_+}{\gamma\Big(\frac{3}{2},\frac{r_+^2}{4\theta}\Big)}\,dr_+,
\end{equation}
where
\begin{equation}
\zeta=1+\frac{a\, \alpha}{\sqrt{1-\frac{a^2\,\alpha^2}{2}}}.
\end{equation}
We recover the entropy of the rotating of the black string in the limit $r_+/\sqrt{\theta}\to \infty$ as
\begin{equation}
S_+=\frac{\pi r_+^2}{2\zeta}
\end{equation}
Finally, we analyze the thermodynamical stability of the NC geometry inspired rotating black string. The specific heat is given by Eq. (\ref{eqcp}), substituting (\ref{eqmr}) and (\ref{eqtr}) into (\ref{eqcp}), we find the heat capacity of the NC geometry inspired rotating black string as
\begin{eqnarray}
C_+=\frac{\pi^{3/2} r_{+}^2\Bigg[3-\frac{r_+\gamma'\left(\frac{3}{2},\,\frac{r_{+}^2}{4\,\theta}\right)}{\gamma\left(\frac{3}{2},\,\frac{r_{+}^2}{4\,\theta}\right)}\Bigg]}{2\zeta\Bigg[3\gamma\left(\frac{3}{2},\,\frac{r_{+}^2}{4\,\theta}\right) -2r_+\left[\gamma'\left(\frac{3}{2},\,\frac{r_{+}^2}{4\,\theta}\right)+r_+\gamma''\left(\frac{3}{2},\,\frac{r_{+}^2}{4\,\theta}\right)\right]+r_+^2\frac{\gamma'\left(\frac{3}{2},\,\frac{r_{+}^2}{4\,\theta}\right)}{\gamma\left(\frac{3}{2},\,\frac{r_{+}^2}{4\,\theta}\right)}\Bigg]},
\label{cap2}
\end{eqnarray}
again the large radius limit leads to
\begin{equation}
C_+= \frac{\pi r_+^2}{\zeta},
\label{shr}
\end{equation}
\begin{figure}[h]
\begin{tabular}{c c c c}
\includegraphics[width=0.55\linewidth]{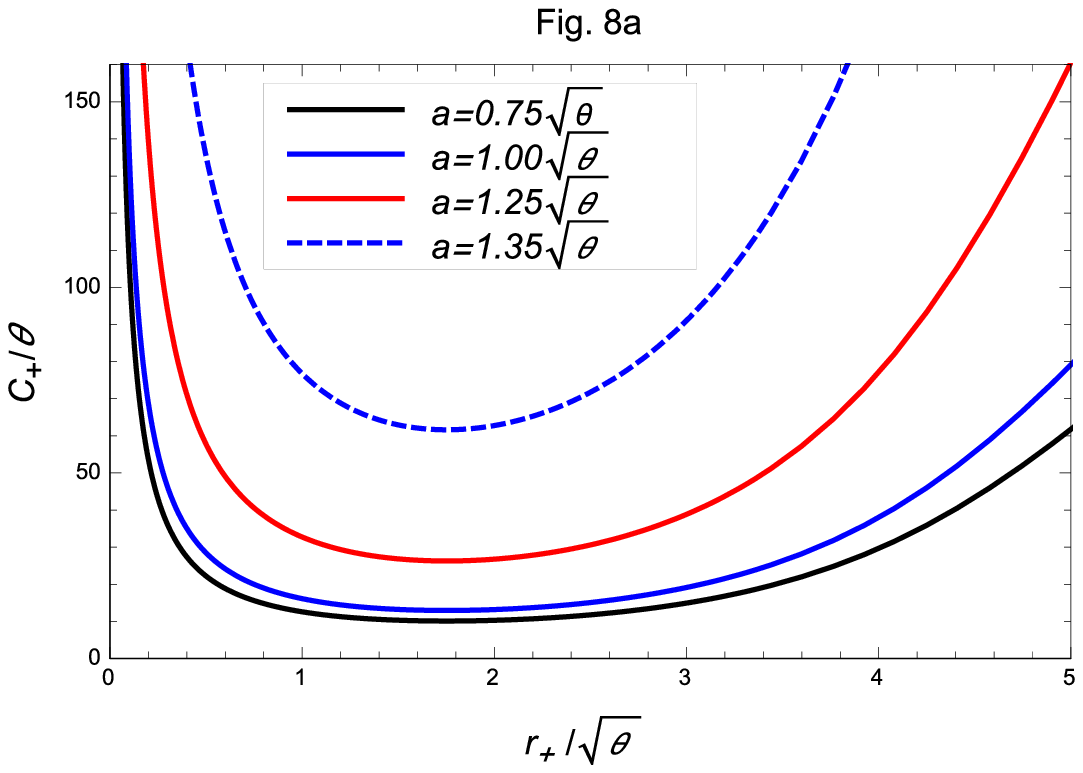}
\includegraphics[width=0.55\linewidth]{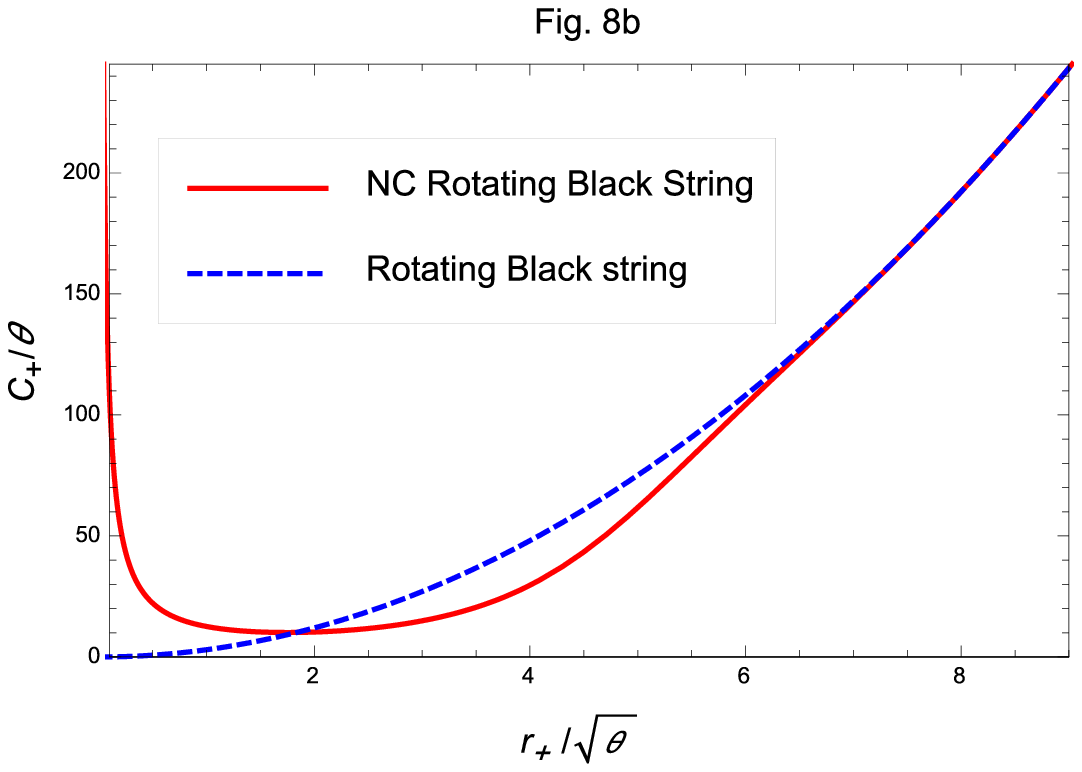}
\end{tabular}
\caption{\label{CS31} Plot of (a) the specific heat ($C_+/\theta$) vs  horizon radius $r_+/\sqrt{\theta}$ with different value of rotation parameter $a/\sqrt{\theta}$ (left) and (b) specific heat vs ($r_+/\sqrt{\theta}$), the solid and dashed lines show the heat capacity of NC geometry inspired rotating black string and rotating black string with fixed value of rotation parameter $a=0.75\sqrt{\theta}$.}
\end{figure}
\noindent which is the specific heat of commutative rotating black string (\ref{metric3}). 

\noindent In the standard rotating black string solution heat capacity vanishes as horizon radius shrinks to zero. But in case of NC geometry inspired rotating black string the heat capacity does not vanish even if the horizon radius is zero (cf. Fig. \ref{CS31}b). We observed that the heat capacity has single stable phase in the interval $0<a^2\alpha^2<2/3$ and it shifts toward the divergences with increase in the value of rotation parameter $a$ (cf. Fig. \ref{CS31}a). As $a^2\alpha^2$ approaches to 2/3, the heat capacity diverges. In the region $r_+\simeq\sqrt{\theta}$ the specific heat ($C_+$) deviates from the standard result (\ref{shr}), and coincides at $r_+=8.91\sqrt{\theta}$ with fixed value of rotation parameter $a=0.75\sqrt{\theta}$ (cf. Fig. \ref{CS31}b).
\section{Conclusions}
Motivated by a string theory arguments \cite{snyder}, NC geometry inspired black hole spacetime has been explored extensively. In this paper, we have obtained NC geometry inspired cylindrically symmetric rotating black string solutions at a short distance, and thus Lemos black string solution is recovered in the large distance limit. It has been shown by Nicolini et. al \cite{nicoli06} that noncommutatitivity can be introduced in general relativity by modifying the energy-momentum tensor. The resulting energy-momentum tensor has description of perfect fluid form with non-zero pressure and with the energy density distribution of diffused source. After solving  Einstein's field euations with this source, one can find the metric funtion of NC geometry inspired balck holes. Similiarly, we have applied this recipe to obtain the NC geometry inspired balck string (including its rotating counterpart) and discuss it's behaviours in the region $r\approx\sqrt{\theta}$. Interestingly, it turns out that NC cures the divergences appearing in various form in the classical general relativity.

The black string metrices are governed by an anisotropic type of matter, whose non-vanishing pressure prevents matter collapsing into a singularity. Our black string solution contains their general relativity counterpart in the limit $r/\sqrt{\theta}\to \infty$. The construction for the existence of extreme black string is derived. Despite of complexity due to NC geometry, we have obtained the exact expression for the thermodynamic quantities associatd with the NC geometry inspired black string, such as the Hawking temperature, the entropy and the heat capcity. In particular, the heat capacity is analysed in detail and shown that heat capacity is always positive at all radius. Both the NC geometry inspired rotating and non-rotating black strins have two horizon corresponding to certain values mass parameter, but one of the horizons (the inner or Cauchy horizon) has very small value($\approx 10^{-4}\sqrt{\theta}$) and therefore, is not visibile in the graphs. The results presented here are the generalization of the previous discussions of Lemos black string in general relativity in a more general setting, and the possibilit in by of a further generalization of these results to higher dimensional BTZ like solution is an interesting problem for future research.
\begin{acknowledgements}
Dharm Veer Singh acknowledges the University Grant Commission, India, for financial support through the D. S. Kothari Post Doctoral Fellowship (Grant No.: BSR/2015-16/PH/0014). S.G.G. would like to thanks SERB-DST Research Project Grant No. SB/S2/HEP-008/2014 and DST INDO-SA bilateral project DST/INT/South Africa/P-06/2016 and also to IUCAA, Pune for the hospitality while this work was being done. 
\end{acknowledgements}

\end{document}